\documentclass[
    ,final            
  ]
  {aipproc}

\layoutstyle{6x9}

\begin{document}

\title{The mass of charmonium in nuclear matter}

\author{Su Houng Lee}{address={Department of Physics and Institute of Physics and Applied Physics,
Yonsei University, Seoul 120-749, Korea} }

\begin{abstract}
The masses of charmonium states immersed in nuclear matter are
calculated in LO QCD and in QCD sum rules.   While the mass shift
for $J/\psi$ are found to be less than -10 MeV, those for the
$\chi_{0,1,2}$ and $\psi(3686)$ and $\psi(3770)$ are found to be
more than -40 MeV.  We investigate the feasibility of observing
such mass shifts in the future accelerator project at GSI.
\end{abstract}

\maketitle


\section{Introduction}
Understanding hadron mass changes in nuclear medium and/or at
finite temperature can provide valuable information about the QCD
vacuum\cite{brown,HL92,mosel}.  While the mass shifts for hadrons
made of light quarks are sensitive to the restoration of the
spontaneously broken chiral symmetry breaking\cite{HL92,shlee,
Hayashigaki00,Tsushima99}, those for the the heavy quark systems
are sensitive to the changes of the non-perturbative gluon fields
in nuclear matter. For the $J/\psi$, which consists of a charm and
anticharm quark pair, both the QCD sum rules analysis
\cite{Kli99,KL01} and the LO perturbative QCD calculation
\cite{Peskin79,Luk92} show that its mass is reduced slightly in
the nuclear matter mainly due to the reduction of the gluon
condensate ($\langle \frac{\alpha_s}{\pi} G^2 \rangle$) in nuclear
matter, which is expected to decrease by 6\% at normal nuclear
matter density.   However, the changes are much larger for excited
charmonium states, due mainly to larger color dipole size of these
excited states.

In this report, we summarize the expected mass shift for
charmonium states in nuclear matter and study the feasibility of
observing such mass shift in the future accelerator project at
GSI\cite{GSI-future}.

The lowest dimensional QCD operators that characterizes the non
perturbative nature of the QCD vacuum are the quark and gluon
condensate.  These condensate are estimated to have the following
large non-perturbative expectation values in the vacuum
\cite{SVZ},
\begin{eqnarray}
\langle \frac{\alpha_s}{\pi} F_{\mu \nu}^2 \rangle & \sim & 1. 5~
{\rm GeV/fm}^3,
\nonumber \\
\langle \bar{q} q \rangle & \sim &  2 ~{\rm fm}^{-3}.
\end{eqnarray}
The gluon condensate can be written as the difference between the
magnetic $B^2= F_{ij}^2$ and electric $E^2=\frac{1}{2} F_{0i}^2$
condensate, which respectively contribute to half of the zero
temperature gluon condensate,
\begin{eqnarray}
\langle \frac{\alpha_s}{\pi} B^2 \rangle =-\langle
\frac{\alpha_s}{\pi} E^2 \rangle=\frac{1}{2}\langle
\frac{\alpha_s}{\pi} F_{\mu \nu}^2 \rangle. \label{BE}
\end{eqnarray}
The above relation follows naturally from the Euclidean
formulation of QCD at zero temperature, such as in the lattice
QCD, where the Euclidean  space electric (magnetic) condensate is
defined with a minus (plus) sign relative to its Minkowski space
counterpart\cite{SVZ}.  Due to the symmetry in the time and space
directions in the 4 dimensional Euclidean space, the Euclidean
space electric condensate is expected to have the same expectation
value as the magnetic one\cite{Dig,Lee89}.  Hence the relation
among the Minkowski space condensate in Eq.(\ref{BE}) follows.

At nuclear matter, the non perturbative quark and gluon field
configuration are expected to change appreciably, such that the
average gluon and quark condensate values decrease by 6\% and
30\%, respectively.  These model independent results are obtained
from the linear density approximation and the nucleon expectation
values of the quark and gluon condensate, which are respectively
known from the experimentally measured pi-N sigma term and from
taking the nucleon expectation value of the trace anomaly
relation\cite{SVZ2}.   The electric and magnetic part of the gluon
condensate at nuclear matter can be estimated separately, by using
the twist-2 gluon operator,
\begin{eqnarray}
\langle N(p)| {\cal ST} F_\mu^\alpha F_{\alpha \nu} |N(p) \rangle
= \bigg( p_\mu p_\nu -\frac{1}{4} m_N^2 g_{\mu \nu} \bigg) 2
A_2(g),
\end{eqnarray}
where $A_2(g)$ is the second moment of the gluon distribution in
the nucleon and is around 0.45, when the renormalization scale is
in the of order 1 to 2 GeV.   Using this and the linear density
approximation, we have,
\begin{eqnarray}
\langle \frac{\alpha_s}{\pi} E^2 \rangle_{n.m.}  & = & \bigg(
\frac{4}{9} m_N m_N^0 + \frac{3}{2} m_N^2 \frac{\alpha_s}{\pi}A_2
\bigg) \frac{\rho_{n.m.}}{2m_N}, \nonumber \\
\langle \frac{\alpha_s}{\pi} B^2 \rangle_{n.m.}  & = & -\bigg(
\frac{4}{9} m_N m_N^0 - \frac{3}{2} m_N^2 \frac{\alpha_s}{\pi}A_2
\bigg) \frac{\rho_{n.m.}}{2m_N}. \label{BE2}
\end{eqnarray}
Here, $m_N^0\sim 0.75 $ GeV is the mass of the nucleon in the
chiral limit\cite{BM96}, which comes from taking the nucleon
expectation value of the trace anomaly relation
$T_\mu^\mu=-\frac{9}{8} \frac{\alpha_s}{\pi}F^2_{\mu \nu}$. As can
be seen from Eq.(\ref{BE2}), due to the additional factor of
$\frac{\alpha_s}{\pi}$ in the second terms, the changes are
dominated by the contribution from the first terms.

\section{Charmonium mass shift from QCD}

The mass shift of charmonium states in nuclear medium can be
evaluated in the perturbative QCD when the charm quark mass is
large, i.e., $m_c \to \infty$.  In this limit,
one can perform a systematic operator product expansion (OPE) of
the charm quark-antiquark current-current correlation function
between the heavy bound states by taking the separation scale
($\mu$) to be the binding energy of the charmonium
\cite{Peskin79,BP79,OKL02}. The forward scattering matrix element
of the charm quark bound state with a nucleon then has the
following form:
\begin{eqnarray}
T(q^2=m_\psi^2)= \sum_n \frac{C_n}{(\mu)^n} \langle {\cal O}_n
\rangle_N. \label{ope}
\end{eqnarray}
Here, $C_n$ is the Wilson coefficient evaluated with the charm
quark bound state wave function and $\langle {\cal O}_n \rangle_N$
is the nucleon expectation value of local operators of dimension
$n$.

For heavy quark systems, there are only two independent lowest
dimension operators; the gluon condensate ($\langle
\frac{\alpha_s}{\pi} G^2 \rangle$) and the condensate of twist-2
gluon operator multiplied by $\alpha_s$ ($\langle \frac{\alpha_s}
{\pi}G_{\alpha\mu} G^{\alpha}_\nu \rangle$). These operators can
be rewritten in terms of the color electric and magnetic fields:
$\langle \frac{\alpha_s}{\pi} E^2 \rangle$ and  $ \langle
\frac{\alpha_s}{\pi} B^2 \rangle$.  Since the Wilson coefficient
for $\langle \frac{\alpha_s}{\pi} B^2 \rangle$ vanishes in the
non-relativistic limit, the only contribution is thus proportional
to $\langle \frac{\alpha_s}{\pi} E^2 \rangle$, similar to the
usual second-order Stark effect. We shall thus calculate the mass
shift of charmonium states due to change of the gluon condensate
in nuclear medium by the QCD second-order Stark effect
\cite{Luk92}.

The mass shift of charmonium states to leading order in density is
obtained by multiplying the leading term in Eq.(\ref{ope}), by the
nuclear density $\rho_N$. This gives,
\begin{eqnarray}
\Delta m_{\psi} (\epsilon) & = &  -\frac{1}{9} \int d k^2 \bigg|
{\partial \psi(k) \over  \partial {\bf k}} \bigg|^2 {k
\over k^2/m_c+ \epsilon } \nonumber \\
&& \times  \bigg\langle \frac{\alpha_s}{\pi} E^2 \bigg\rangle_N
\cdot \frac{\rho_N}{2 m_N}. \label{stark}
\end{eqnarray}
In the above, $m_N$ and $\rho_N$ are the nucleon mass and the
nuclear density, respectively; $\langle \frac{\alpha_s}{\pi} E^2
\rangle_N\sim 0.5$ GeV$^2$ is the nucleon expectation value of the
color electric field obtained from Eq.(\ref{BE2}) and $\epsilon
=2m_c-m_\psi$. In Ref.\cite{Peskin79}, the LO mass shift formula
was derived in the large charm quark mass limit. As a result, the
wave function $\psi(k)$ is Coulombic and the mass shift is
expressed in terms of the Bohr radius $a_0$ and the binding energy
$\epsilon_0=2m_c-m_{J/\psi}$. This might be a good approximation
for $J/\psi$ but is not realistic for the excited charmonium
states as Eq.(\ref{stark}) involves the derivative of the wave
function, which measures the dipole size of the system. We have
thus rewritten in the above the LO formula for charmonium mass
shift in terms of the QCD parameters $\alpha_s=0.84$ and
$m_c=1.95$, which are fixed by the energy splitting between
$J/\psi$ and $\psi(3686)$ in free space\cite{Peskin79}.
Furthermore, we take wave functions of the charmonium state to be
Gaussian with the oscillator constant $\beta$ determined by their
squared radii $\langle r^2 \rangle=$ $0.47^2$, $0.74^2$, $0.96^2$,
and 1 fm$^2$ for $J/\psi$, $\chi_{0,1,2}$,  $\psi(3686)$, and
$\psi(3770)$, respectively, as obtained from the potential models
\cite{cornel}. This gives $\beta=$ 0.52, 0.43, 0.39, and 0.37 GeV
if we assume that these charmonium states are in the $1S$, $1P$,
$2S$, and $1D$ states, respectively. Using these parameters, we
find that the mass shifts at normal nuclear matter density
obtained from the LO QCD formula Eq.(\ref{stark}) are -8, -40,
-100, and -140 MeV for $J/\psi$, $\chi_{0,1,2}$, $\psi(3686)$, and
$\psi(3770)$, respectively\cite{LK03}.

Although the higher twist effects on the charmonium masses are
expected to be nontrivial, the result for $J/\psi$ is consistent
with those from other non-perturbative QCD studies, such as the
QCD sum rules \cite{Kli99,KL01} and the effective potential model
\cite{Bro90,Was91,Navarra95}, which are all based on the dipole
interactions between quarks in the charmonium and those in the
nuclear matter.  The QCD sum rule results can also be applied for
the $\chi_{0,1,2}$ states, and the results from the leading order
gluon condensate is summarized in Table. \ref{t1}.

Higher twist effects can be estimated in some calculations.  For
QCD sum rules for $J/\psi$, the corrections coming from dimension
6 operators are less than 30\% of the leading order
results\cite{KL01}.   The contributions from the $D \bar{D}$ meson
loops in the $\psi(3686)$ and $\psi(3770)$ are also found to be
less than 30\% of the LO QCD result\cite{LK03}.

All the results are summarized in table \ref{t1}.

\begin{table}[ht]
\caption{Charmonium Mass shift in nuclear matter in MeV}
\label{t1}
\begin{tabular}{ccccc}
\hline
Charmonium & $J^{PC}$ & QCD 2nd order Stark Effect & QCD sum rules & Effects of $D \bar{D}$ loop \\
\hline
$\eta_c$ & $0^{-+}$ & - 8 MeV  & -5 MeV & No effect \\
$J/\psi$ & $1^{--}$ & -8 MeV  & -7 MeV & < 2 MeV  \\
$\chi_{0,1,2}$ & $0,1,2^{++}$ & - 40 MeV & -60 MeV & No effect on $\chi_1$  \\
$\psi(3686)$ & $1^{--}$ & -100 MeV & & < 30 MeV \\
$\psi(3770)$ & $1^{--}$ & -140 MeV & & < 40 MeV \\ \hline
\end{tabular}
\end{table}

\section{Observability}

Since the mass shift of the heavy quark system reflects the
changes of the Gluon field configuration in the vacuum, it would
be interesting to observe such effects in experiment.

Consider an anti-proton with incoming four momentum $(\omega,0,k)$
annihilating a proton at rest $(m_N,0,0)$ and creating a
charmonium moving with velocity $v$.  The required incoming
momentum $k$ to create a charmonium state are summarized in Table
I.

\begin{table}[ht]
\caption{Required momentum to create Charmonium with outgoing
velocity $v$} \label{t0}
\begin{tabular}{cccccccc}
\hline
 & $\eta_c$ & $J/\psi$ & $\chi_0$ & $\chi_1$ & $\chi_2$ & $\psi(3686)$ & $\psi(3770)$  \\
\hline
$k$ (GeV/c) & 3.7 & 4.1 & 5.2 & 5.5 & 5.7 & 6.2 & 6.5  \\
$\omega$ (GeV) & 3.8 & 4.2 & 5.3 & 5.6 & 5.8 & 6.3 & 6.6 \\
$v$ (c)& 0.78 & 0.8 & 0.83 & 0.84 & 0.85 & 0.86 & 0.87   \\ \hline
\end{tabular}
\end{table}

Hence,  the required incoming energy of the anti-proton to produce
the charmonium state range from 4 to 6 GeV.  In these energy
region, the absorption cross section $\sigma_{\bar{p} -p} \sim 50
$ mb. Hence, the anti-proton would be absorbed after travelling
less than 1 fm in the nuclear matter.  Moreover, once a charmonium
state is created, their speed would be less than 0.9 c, which
means that it will have to travel more than 10 fm/c to pass the
diameter of a nucleus of A=125.  Hence, considering the increased
width of charmonium due to  nuclear absorption\cite{martins}, the
charmonium is expected to decay inside the nucleus.

The cross section for the production of charmonium states and its
subsequent decay into dileptons or $J/\psi+\gamma$ states are
given by the following Breit-Wigner formula

\begin{eqnarray}
\sigma_{BW}(E)= {2J+1 \over (2s_1+1) (2s_2+1) } \frac{\pi}{k^2}
{B_{in} B_{out} \Gamma_{Total}^2 \over
(E-E_R)^2+\Gamma^2_{Total-medium}/4},
\end{eqnarray}
where $k$ is the c.m. momentum, E is the c.m. energy, $B_{in}$ and
$B_{out}$ are the branching fractions of the resonance into the
entrance and exit channels.  The $2s+1$ are the spin
multiplicities of the incident spin states and $J$ the spin of the
charmonium.  Also
$\Gamma_{Total-medium}=\Gamma_{Total}+\Gamma_{medium}$.

If we substitute the medium mass shift and increase in width (due
mainly to collision broadening), the cross sections are in the
order of one to few hundred pbarn

The expected luminosity at the anti proton project at GSI is $2
\times 10^{32} {\rm cm}^{-2} s^{-1}$. Therefore  if the cross
section is 1pb, it would corresponds to about 17 events per day.

\begin{table}[ht]
\caption{Measurable decay channel and expected event rate at GSI
future accelerator.}
\label{t2}
\begin{tabular}{ccccc} \hline
Charmonium  & $\Gamma_{Total}+\Gamma_{medium}$  &
Final state & cross-section to final state & events per day    \\
\hline\hline
 $ J/\psi(3097)$  &  87 KeV+ 20 MeV   & $e^+ +e^-$
 &  6pb & 100  \\
 $\psi(3686)$  &  300 KeV+ 20 MeV & $e^+ +e^-$ &
 0.6 pb  & 10
 \\ $\psi(3770)$  &  23.6 MeV +20 MeV  & $e^+ +e^-$ &
 1 pb &  17   \\
\hline $\chi_{c0}(3417)$ &  16.2 MeV+ 20 MeV  & $J/\psi+\gamma$ &
200 pb
 & 3400  \\
 $\chi_{c1}(3510)$  & 0.92 MeV+20 MeV  &$J/\psi+\gamma$
 & 80 pb & 1360 \\ $\chi_{c2}(3556)$  & 2.08MeV+20 MeV &$J/\psi+\gamma$
  & 350 pb   & 5950  \\
\hline
\end{tabular}

\end{table}

Hence, the mass shift will be observable in the anti-proton
project at the future accelerator facility at GSI.


\begin{theacknowledgments}

I would like to thank C.M. Ko and A.  Gillitzer for useful
discussions.

\end{theacknowledgments}


\bibliographystyle{aipprocl} 


\IfFileExists{\jobname.bbl}{}
 {\typeout{}
  \typeout{******************************************}
  \typeout{** Please run "bibtex \jobname" to optain}
  \typeout{** the bibliography and then re-run LaTeX}
  \typeout{** twice to fix the references!}
  \typeout{******************************************}
  \typeout{}
 }

\end{document}